\documentclass[preprint,superscriptaddress,amssymb,amsmath,floatfix,citeautoscript]{revtex4-1}
\usepackage[pdftex]{graphicx}
\usepackage{amsmath,amssymb}
\usepackage{color}
\usepackage{natbib}

\begin{document}

This manuscript has been authored by UT-Battelle, LLC under Contract No. DE-AC05-00OR22725 with the U.S. Department of Energy. The United States Government retains and the publisher, by accepting the article for publication, acknowledges that the United States Government retains a non-exclusive, paid-up, irrevocable, world-wide license to publish or reproduce the published form of this manuscript, or allow others to do so, for United States Government purposes. The Department of Energy will provide public access to these results of federally sponsored research in accordance with the DOE Public Access Plan (http://energy.gov/downloads/doe-public-access-plan).

\newpage

\title{Space-time dependent thermal conductivity in nonlocal thermal transport}

\author{Chengyun Hua\footnote{\text{
To whom correspondence should be addressed. E-mail: huac@ornl.gov}}}
\affiliation{Materials Science and Technology Division, Oak Ridge National Laboratory, Oak Ridge, TN 37831, USA}
\author{Lucas Lindsay}
\affiliation{Materials Science and Technology Division, Oak Ridge National Laboratory, Oak Ridge, TN 37831, USA}

\date{\today}

\begin{abstract}

Nonlocal thermal transport is generally described by the Peierls-Boltzmann transport equation (PBE). However, solving the PBE for a general space-time dependent problem remains a challenging task due to the high dimensionality of the integro-differential equation. In this work, we present a direct solution to the space-time dependent PBE with a linearized collision matrix using an eigendecomposition method. We show that there exists a generalized Fourier type relation that links heat flux to the local temperature, and this constitutive relation defines a thermal conductivity that depends on both time and space. Combining this approach with ab initio calculations of phonon properties, we demonstrate that the space-time dependent thermal conductivity gives rise to an oscillatory response in temperature in a transient grating geometry in high thermal conductivity materials. The present solution method allows us to extend the reach of our computational capability for heat conduction to space-time dependent nondiffusive transport regimes. This capability will not only enable a more accurate interpretation of thermal measurements that observe nonlocal thermal transport but also enhance our physical understanding of nonlocal thermal transport in high thermal conductivity materials that are promising candidates for nanoscale thermal management applications.

\end{abstract}

\maketitle

\section{Introduction}
Nonlocal thermal transport, which occurs when a temperature gradient exists over a length scale comparable to or smaller than the mean free paths (MFPs) of the heat carriers, is a subject of considerable interest for basic science and thermal management technologies. Recent experiments have demonstrated ultrahigh thermal conductivity in Boron Arsenide\cite{kang_experimental_2018,li_high_2018, tian_unusual_2018} and isotope-enriched cubic Boron Nitride\cite{chen_ultrahigh_2020}, two viable candidates for novel thermal management applications including substrates for high-power electronics. Nonlocal thermal transport can be easily observed in these ultrahigh thermal conductivity dielectric crystals due to a large fraction of long MFP phonons, the dominant heat carriers in these materials. An accurate mathematical description of nonlocal thermal transport is critical to developing further insights into associated effects toward utilization in practical applications.

Nonlocal phonon transport in crystals is generally described by the Peierls-Boltzmann equation (PBE)\cite{Peierls1929},
\begin{equation}\label{eq:PBTE}
\frac{\partial f_{\mu}}{\partial t}+\mathbf{v_{\mu}}\cdot \nabla f_{\mu} = -\frac{\partial f_{\mu}}{\partial t}\biggm|_{scattering},
\end{equation}
which describes the dynamics of the out-of-equilibrium occupation function $f_{\mu}$ at position $\mathbf{x}$ and time $t$, for all possible phonon states $\mu$ ($\mu \equiv (\mathbf{q},s)$, where $\mathbf{q}$ is the phonon wavevector and $s$ is the phonon polarization). In this equation, $\mathbf{v_{\mu}}$ is the phonon group velocity. Solving the PBE for a general space-time dependent problem remains a challenging task due to the high dimensionality of the integro-differential equation. 

Thus, most prior works have determined solutions of the PBE under various assumptions. One widely used assumption is the single-mode relaxation time approximation (RTA), where each phonon mode relaxes towards thermal equilibrium at a characteristic relaxation rate independent of the other phonons. The RTA has been used to investigate nonlocal transport in an infinite domain\cite{Mahan_1988,hua_analytical_2014,allen_temperature_2018,Collins2013APL}, a finite one-dimensional slab\cite{Hua_semi_analytical_2015, Koh_2014}, and experimental configurations such as transient grating\cite{Ramu2014,Collins2013APL} and thermoreflectance\cite{Regner2014,zeng_disparate_2014, Vermeersch_2015a,Vermeersch_2015b} measurements. An efficient Monte Carlo scheme was used to solve the PBE under the RTA for complicated geometries involving multiple boundaries\cite{Peraud:2011,Peraud:2012,hua_importance_2014}. However, the RTA introduces difficulties in defining pseudo local temperature as noted by Peraud\cite{Peraud:2011,Peraud:2012}. No satisfactory explanation has thus far been given regarding the need for an additional "temperature definition". Moreover, first principles calculations have demonstrated that the RTA fails to adequately describe thermal transport in materials with weak intrinsic thermal resistance\cite{ward_intrinsic_2010, lindsay_first_2016}. In short, the RTA assumption is not appropriate for materials with ultrahigh thermal conductivity, such as diamond or cubic Boron Nitride, which also have phonons with long MFPs and presumably important nonlocal transport effects.  

Some efforts have attempted to solve the PBE with a linearized collision operator. Guyer and Krumhansl\cite{guyer_solution_1966} first performed a linear response analysis of the PBE, deriving a space-time-dependent thermal conductivity by assuming the Normal scattering rates were much larger than Umklapp scattering rates. They applied their solution to develop a phenomenological coupling between phonons and elastic dilatational fields caused by lattice anharmonicity. Hardy and coworkers reported a rigorous quantum-mechanical formulation of the theory of lattice thermal conductivity using a perturbation method that included both anharmonic forces and lattice imperfections\cite{hardy_energy-flux_1963,hardy_perturbation_1965, hardy_lowestorder_1965}. This quantum treatment of lattice dynamics was then verified both theoretically and experimentally demonstrating the presence of Poiseuille flow and second sound in a phonon gas at low temperatures when Umklapp processes can be neglected\cite{sussmann_thermal_1963, guyer_thermal_1966, hardy_phonon_1970, jackson_thermal_1971,beck_phonon_1974}. The variational principle has also been used to solve the PBE with Umklapp scattering incorporated\cite{hamilton_variational_1969, srivastava_derivation_1976}. Levinson developed a nonlocal diffusion theory of thermal conductivity from a solution of the PBE with three-phonon scattering in the low frequency limit\cite{levinson_nonlocal_1980}.

More recently, lattice thermal conductivity has been computed from first principles by imposing a constant temperature gradient and using an iterative method\cite{ward_ab_2009,broido_lattice_2005,li_shengbte:_2014,carrete_almabte_2017,omini_iterative_1995} or a variational approach\cite{fugallo_ab_2013} to solve the PBE. Typical first principles PBE calculations consider only a linear spatial temperature profile without time variation. Chaput presented a direct solution to the time-dependent PBE imposed with a linear temperature profile by computing the eigenvalues and eigenvectors of a symmetrized matrix of reduced dimensions\cite{chaput_direct_2013}.  Dynamical thermal conductivity with a terahertz temporal frequency was calculated for the first time. 

Li and Lee\cite{li_role_2018} studied the role of hydrodynamic viscosity on phonon transport in suspended graphene using the Monte Carlo solution of the PBE with an ab initio full three-phonon scattering matrix first introduced by Landon and Hadjiconstantinou\cite{landon_deviational_2014}. The peculiar thermal conductivity dependence on sample width was explained with a phonon viscous damping effect in the hydrodynamic regime. 

Cepellotti and Marzari\cite{cepellotti_thermal_2016} introduced the concept of a "relaxon", an eigenstate of the symmetrized scattering operator of the PBE, first used by Guyer \emph{et. al.}\cite{guyer_solution_1966} and Hardy\cite{hardy_phonon_1970} in their studies of second sound\cite{guyer_thermal_1966,hardy_phonon_1970}. 
They applied this treatment to solve steady-state problems in two-dimensional systems with a constant temperature gradient\cite{cepellotti_boltzmann_2017}. They also showed that the derived relaxons had well-defined parity, with odd relaxons governing  thermal conductivity\cite{cepellotti_thermal_2016} and even relaxons contributing to thermal viscosity\cite{simoncelli_generalization_2020}. These quantities together give a macroscopic description of heat transport in the hydrodynamic regime\cite{simoncelli_generalization_2020}.

However, these previous efforts either require expansive numerical simulations or contain various assumptions that may conceal important insights of nondiffusive thermal transport behaviors. In this work, we use a similar eigendecomposition method first proposed by Guyer\cite{guyer_solution_1966} and Hardy\cite{hardy_energy-flux_1963,hardy_perturbation_1965, hardy_lowestorder_1965} to solve the PBE with a generalized linearized collision matrix. The only assumption in this method is the linearization of the collision matrix.  We will demonstrate that a generalized Fourier's law, similar to that derived in Ref. \cite{hua_generalized_2019}, also exists in this linear regime. We apply this theoretical construct to examine spatial and temporal thermal transport in diamond, Si, Ge, and cubic BN from first principles. 

\section{Governing Equations}
Starting with Eq.~(\ref{eq:PBTE}), the collision operator can be linearized around the global equilibrium Bose-Einstein distribution $f^0_{\mu}(T_0) = (\text{exp}(\hbar\omega_{\mu}/(k_BT_0))-1)^{-1}$, where $\omega_{\mu}$ is the phonon frequency, $k_B$ is the Boltzmann constant, and $T_0$ is the equilibrium temperature. The linearized BTE can be written into the following form\cite{chaput_direct_2013,srivastava_physics_1990}, 
\begin{equation}\label{eq:BTE_raw}
\frac{\partial \Delta f_{\mu}}{\partial t}+\mathbf{v_{\mu}}\cdot \nabla (\Delta f_{\mu} )= -\frac{1}{\nu}\sum_{\mu'}\Omega_{\mu\mu'}\Delta f_{\mu'}\frac{\text{sinh}(\frac{\hbar\omega_{\mu'}}{2k_BT_0})}{\text{sinh}(\frac{\hbar\omega_{\mu}}{2k_BT_0})},
\end{equation}
where $\Delta f_{\mu} = f_{\mu}-f^0_{\mu}(T_0)$, $\nu$ is a normalized volume, and $\Omega_{\mu,\mu'}$ is the linear phonon scattering operator. The right hand side of Eq.~(\ref{eq:BTE_raw}) describes scattering as a linear operator represented by the action of the matrix $\Omega_{\mu\mu'}$ on $\Delta f_{\mu'}\text{sinh}(\hbar\omega_{\mu'}/(2k_BT_0))$. This linearization of the scattering operator has been used in most studies of thermal transport and holds for small deviations from thermal equilibrium.\cite{Ziman1960,chaput_direct_2013,cepellotti_thermal_2016,ward_ab_2009,broido_lattice_2005,li_shengbte:_2014} The scattering matrix appearing in Eq.~(\ref{eq:BTE_raw}) is in its most general form and describes all possible mechanisms by which a phonon excitation can be transferred from a state $\mu$ to a state $\mu'$ regardless of interaction mechanism. The matrix operator representing three-phonon interactions is given in the appendix.

The matrix $\Omega$ has four key features: 1) it is real and symmetric, \emph{i.e.} $\Omega_{\mu\mu'} = \Omega_{\mu'\mu}$; 2) it is an even function of $\mu$, \emph{i.e.} $\Omega_{-\mu-\mu'} =\Omega_{\mu\mu'}$; 3) it is positive semi-definite, \emph{i.e.} $|\Omega_{\mu\mu'}|\geq 0$; 4) it is summational invariant, \emph{i.e.} $\sum_{\mu} \hbar\omega_{\mu}\text{sinh}^{-1}(\hbar\omega_{\mu}/(2k_BT_0))\Omega_{\mu\mu'} =0$. Therefore, when multiplying Eq.~(\ref{eq:BTE_raw}) with $\hbar\omega_{\mu}$ and then integrating over $\mu$ in the Brillouin zone, the equation of energy conservation is recovered as
\begin{equation}\label{eq:EnergyConservation}
\frac{\partial \Delta E}{\partial t}+\nabla \cdot \mathbf{J} = 0,
\end{equation}
where $\Delta E= \nu^{-1}\sum_{\mu}\hbar\omega_{\mu}\Delta f_{\mu}$ is the deviational energy and $\mathbf{J} =\nu^{-1} \sum_{\mu}\hbar\omega_{\mu}\mathbf{v_{\mu}}\Delta f_{\mu}$ is the heat flux in and out of the control volume. 

The goal of the following mathematical treatment is to find the appropriate transform matrix to rotate the highly coupled linear system represented by Eq.~(\ref{eq:BTE_raw}) into a set of decoupled linear equations. To do this, we use a spectral decomposition method, first used by Guyer and Krumhansl\cite{guyer_solution_1966}. First, we perform a change of variables by defining
\begin{equation}
n_{\mu} = \Delta f_{\mu}\text{sinh}\left(\frac{\hbar\omega_{\mu}}{2k_BT_0}\right).
\end{equation}
Then Eq.~(\ref{eq:BTE_raw}) becomes
\begin{equation}\label{eq:BTE_linearized}
\frac{\partial n_{\mu}}{\partial t}+\mathbf{v_{\mu}}\cdot \nabla n_{\mu} = -\frac{1}{\nu}\sum_{\mu'}\Omega_{\mu\mu'}n_{\mu}.
\end{equation}

Due to the above mentioned properties of the matrix $\Omega$, we can deduce that there exists a complete set of eigenvectors such that 
\begin{equation}
\frac{1}{\nu}\sum_{\mu'}\Omega_{\mu\mu'}\theta^{\alpha}_{\mu'} = \frac{1}{\tau_\alpha}\theta^\alpha_\mu,
\end{equation}
where $ \tau_\alpha^{-1}$ is the eigenvalue (the lifetime of relaxons introduced by Cepellotti and Marzari \cite{cepellotti_thermal_2016})  and $\alpha$ ($\alpha = $ 0, 1, 2, 3 ...$N$, where $N$ is the dimension of the collision matrix) is the eigenvalue index of matrix $\Omega$. The orthonormal condition and the scalar product are then defined as 
\begin{equation}
\frac{1}{\nu}\sum_{\mu}\theta^{\alpha}_{\mu}\theta^{\alpha'}_{\mu} = \langle \alpha|\alpha' \rangle = \delta_{\alpha\alpha'},
\end{equation}
and
\begin{equation}
 \langle f|g \rangle=\frac{1}{\nu}\sum_{\mu}f_{\mu}g_{\mu} = \langle g|f \rangle.
\end{equation}
Since $\Omega$ is real and symmetric, all of its eigenvectors must be real. Since $\Omega$ is an even function of $\mu$, its eigenvectors can be chosen to be either even or odd, \emph{i.e.} $\theta^\alpha_\mu = \pm \theta^\alpha_{-\mu}$. Because of its positive semi-definiteness, one can show that its eigenvalues are non-negative, \emph{i.e.} $\tau_\alpha \geq 0$ $\forall \alpha$, and only one eigenvalue is necessarily zero, which is labeled as $\alpha = 0$. The associated normalized eigenvector is 
\begin{equation}
\theta^{(0)}_{\mu} = \frac{1}{\sqrt{4k_BT^2_0C_0}}\frac{\hbar\omega_{\mu}}{\text{sinh}\left(\frac{\hbar\omega_{\mu}}{2k_BT_0}\right)},
\end{equation}
where $C_0$ is the volumetric heat capacity given by
\begin{equation}\label{eq:heatcapacity}
C_0 = \frac{1}{\nu}\sum_{\mu}\hbar\omega_{\mu}\frac{\partial f^0_{\mu}}{\partial T}\biggm|_{T_0}.
\end{equation}
Then $n_{\mu}$ is expanded as 
\begin{equation}\label{eq:f_1ex}
n_{\mu} = \sum_{\alpha}g^{\alpha}\theta^{\alpha}_{\mu},
\end{equation}
where $g^{\alpha} = \langle n|\theta^{\alpha} \rangle$.

It follows from Eq.~(\ref{eq:BTE_linearized}) and from the orthogonality and completeness of the eigenvectors that the coefficients $g^{\alpha}$ are determined by the coupled set of equations,
\begin{equation}\label{eq:BTE_coefficient}
\frac{\partial g^{\alpha}}{\partial t}+\sum_{\beta} \langle \alpha|\mathbf{v}|\beta\rangle \cdot \nabla g^{\beta} = -\frac{g^{\alpha}}{\tau^{\alpha}},
\end{equation}
where the matrix elements of the group velocity are
\begin{equation}
\langle \alpha|\mathbf{v}|\beta\rangle = \frac{1}{\nu}\sum_{\mu} \theta^\alpha_{\mu}\mathbf{v_{\mu}}\theta^\beta_{\mu}.
\end{equation}
Since $\mathbf{v_{\mu}} = - \mathbf{v_{-\mu}}$, matrix elements connecting two eigenvectors with the same parity must be zero. 

One can write the total heat flux and total deviational energy in the spectral representation as 
\begin{equation}
\mathbf{J} =\frac{1}{\nu}\sum_{\mu}\hbar\omega_{\mu}\mathbf{v_{\mu}}\Delta f_{\mu} = \sqrt{4k_BT^2_0C_0}\sum_{\beta>0, odd} \langle 0|\mathbf{v}|\beta\rangle g^{\beta},
\end{equation}
and 
\begin{equation} 
\Delta E = \frac{1}{\nu}\sum_{\mu}\hbar\omega_{\mu}\Delta f_{\mu} = \sqrt{4k_BT^2_0C_0}g^{(0)}.
\end{equation} 
Therefore, the zeroth component equation, written as
\begin{equation}
\label{eq:zerothorder}
\frac{\partial g^{(0)}}{\partial t}+\sum_{\beta>0, odd} \langle 0|\mathbf{v}|\beta\rangle \cdot \nabla g^{\beta} = 0,
\end{equation}
is a requirement of energy conservation, equivalent to Eq.~(\ref{eq:EnergyConservation}). Furthermore, $\Delta E$ can be written in terms of the local equilibrium $f^0_{\mu}(\mathbf{x},t)$ defined by a local temperature $T(\mathbf{x},t)$ such that 
\begin{equation}\label{eq:DeviationalEnergy}
\Delta E =\frac{1}{\nu} \sum_{\mu}\hbar\omega_{\mu}[f_{\mu}(\mathbf{x},t)-f^0_{\mu}(T(\mathbf{x},t))]+C_0[T(\mathbf{x},t)-T_0].
\end{equation}
The second term can be rewritten as $C_0\Delta T \langle 0|0\rangle$, where $\Delta T  = T(\mathbf{x},t)-T_0$. Therefore, this is the zeroth component term and the first term in Eq.~(\ref{eq:DeviationalEnergy}) expands over the rest of the eigenvectors. Since $\theta^{(0)}_{\mu}$ is orthogonal to the rest of the eigenvectors, the first term then should be zero such  that $\sum_{\mu}\hbar\omega_{\mu}[f_{\mu}(\mathbf{x},t)-f^0_{\mu}(T(\mathbf{x},t))] = 0$. This determines the formal definition of temperature in this linear regime, 
\begin{equation}\label{eq:temperature}
\Delta E = C_0\Delta T = \sqrt{4k_BT^2_0C_0}g^{(0)}.
\end{equation}
The significance of Eq.~(\ref{eq:temperature}) is that local equilibrium always exists as long as the linearization of the collision operator is valid. Therefore, the macroscopic quantity, temperature, is always well-defined in this regime. 
As Ziman noted\cite{Ziman1960}, to use the definition given by Eq.~(\ref{eq:temperature}), one needs to ensure the subsidiary condition, $\nu^{-1} \sum_{\mu}\hbar\omega_{\mu}[f_{\mu}(\mathbf{x},t)-f^0_{\mu}(T(\mathbf{x},t))] = 0$, is satisfied. In the case of the linearized collision operator, this subsidiary condition is always guaranteed since $\theta^{(0)}_{\mu}$ is orthogonal to $f_{\mu}(\mathbf{x},t)-f^0_{\mu}(T(\mathbf{x},t))$ . However, under the RTA, where the collision operator becomes
\begin{equation}
\frac{\partial f_{\mu}}{\partial t}\biggm|_{scattering} = \frac{f_{\mu}-f^0_{\mu}(\mathbf{x},t)}{\tau^{RTA}_{\mu}}.
\end{equation}
$\theta^{(0)}_{\mu}$ is no longer an eigenvector of the collision matrix, and the subsidiary condition is not necessarily satisfied. The local temperature under the RTA is then defined by satisfying the requirement of energy conservation given by Eq.~(\ref{eq:EnergyConservation}). Previously, such an approach was regarded to define a pseudo temperature\cite{Peraud:2011,Peraud:2012}, but here we argue that it is the true definition of temperature under the RTA rather than Eq.~(\ref{eq:temperature}).

Now, Eq.~(\ref{eq:BTE_coefficient}) can be written into the following system of equations:
\begin{subequations}\label{eq:BTE_coefficient_expanded}
\begin{align}\label{eq:BTE_non-zeroth}
&\frac{\partial g^{\beta}}{\partial t}+\frac{g^{\beta}}{\tau^{\beta}} + \sum_{\alpha>0} \langle\alpha|\mathbf{v}|\beta\rangle \cdot \nabla g^{\alpha} = -\sqrt{\frac{C_0}{4k_BT_0^2}}\langle \beta|\mathbf{v}|0\rangle \cdot \nabla(\Delta T),\\
&\sqrt{\frac{C_0}{4k_BT_0^2}}\frac{\partial \Delta T}{\partial t}+\sum_{\beta>0} \langle 0|\mathbf{v}|\beta\rangle \cdot \nabla g^{\beta} = 0.\label{eq:BTE_zeroth}
\end{align}
\end{subequations}

\section{Solution}
In general, solving the above system requires numerical discretization in time and space and matrix inversion. However, under some specific boundary conditions, one is able to obtain analytical solutions. In the following text, we will derive an analytical solution in a semi-infinite or infinite domain, where the system is subject to a mode dependent small disturbance $\dot{Q}_{\mu}(\mathbf{x},t)$. This small disturbance is equivalent to the volumetric heat generation rate in a diffusion problem, where $\dot{Q}=\frac{1}{\nu}\sum_{\mu}\hbar\omega_{\mu}\dot{Q}_{\mu}(\mathbf{x},t)$. In a similar way, $\dot{Q}_{\mu}(\mathbf{x},t)$ can be expanded as 
\begin{equation}\label{eq:HeatGeneration}
\dot{Q}_{\mu}(\mathbf{x},t) = \sum_{\alpha}q^{\alpha}\theta^{\alpha}_{\mu},
\end{equation}
where $q^{\alpha} = \langle\dot{Q}_{\mu}|\theta^{\alpha}\rangle$. Adding $q^{\alpha}$ to the right hand side of Eq.~(\ref{eq:BTE_coefficient_expanded}) and Fourier transforming it in time and space, we get
\begin{subequations}\label{eq:BTE_FourierTransformed}
\begin{align}\label{eq:BTE_non-zeroth_FourierTransformed}
&i\eta \tilde{g}^{\beta}+\frac{\tilde{g}^{\beta}}{\tau^{\beta}} + i\sum_{\alpha>0} \langle \alpha|\pmb{\xi}\cdot \mathbf{v}|\beta\rangle \tilde{g}^{\alpha} = -\sqrt{\frac{C_0}{4k_BT_0^2}}\langle \beta|\mathbf{v}|0\rangle\cdot (i \pmb{\xi}\Delta \tilde{T})+\tilde{q}^{\beta},\ (\beta>0)\\
&\sqrt{\frac{C_0}{4k_BT_0^2}}i\eta\Delta T+i\pmb{\xi}\cdot \sum_{\beta>0} \langle 0| \mathbf{v}|\beta\rangle \tilde{g}^{\beta} = \tilde{q}^{(0)},\label{eq:BTE_zeroth_FourierTransformed}
\end{align}
\end{subequations}
where $\tilde{g}^{\beta}$, $\Delta \tilde{T}$, and $\tilde{q}^{\beta}$ are the Fourier transformed functions. $\eta$ and $\pmb{\xi} = (\xi_x,\xi_y,\xi_z)$ are the corresponding temporal and spatial variables in Fourier space. Eq.~(\ref{eq:BTE_non-zeroth_FourierTransformed}) is still a highly coupled system of linear equations. To further decouple it, we first perform a change of variables by defining
\begin{equation}\label{eq:define_h}
\tilde{g}^{\beta} =h^{\beta}\sqrt{\tau^{\beta}}.
\end{equation}  
Then, Eq.~(\ref{eq:BTE_zeroth_FourierTransformed}) becomes
\begin{equation}\label{eq:BTE_h}
i\sum_{\alpha}\Gamma^{\alpha\beta}h^{\alpha}+h^{\beta} = -\sqrt{\tau^{\beta}}\sqrt{\frac{C_0}{4k_BT_0^2}} \langle \beta|\mathbf{v}|0\rangle\cdot (i\pmb{\xi}\Delta \tilde{T})+\sqrt{\tau^{\beta}}\tilde{q}^{\beta},
\end{equation}
where $\Gamma^{\alpha\beta} = \sqrt{\tau^{\alpha}}(\delta^{\alpha\beta}\eta+\langle\alpha|\pmb{\xi}\cdot \mathbf{v}|\beta\rangle)\sqrt{\tau^{\beta}}$ and $\delta^{\alpha\beta}$ is a Kronecker delta function. Since $\Gamma$ is a real and symmetric matrix, there exists a complete set of eigenvectors such that
\begin{equation}
\sum_{\alpha}\Gamma^{\alpha\beta}\Phi^{\alpha}_i = \lambda_i\Phi^{\beta}_i,
 \end{equation}
where $\lambda_i$ is the corresponding eigenvalue. The orthonormal condition is given by 
\begin{equation}\label{eq:h_othornormal}
\sum_{\alpha}\Phi^{\alpha}_i\Phi^{\alpha}_j = \delta_{ij},
\end{equation}
and $h^{\beta}$ can be expanded as
\begin{equation}\label{eq:h_expansion}
h^{\beta} = \sum_{i} d_i \Phi^{\beta}_i
\end{equation}
where $d_i = \sum_{\beta}h^{\beta}\Phi^{\beta}_i$ are unknown coefficients to solve for.  Plugging Eq.~(\ref{eq:h_expansion}) into Eq.~(\ref{eq:BTE_h}) and using the orthonormal condition defined by Eq.~(\ref{eq:h_othornormal}), we get 
\begin{equation}
d_i(i\lambda_i+1) = -\sqrt{\frac{C_0}{4k_BT_0^2}}\sum_{\alpha}\Phi^{\alpha}_i\sqrt{\tau_{\alpha}}\langle \alpha|\mathbf{v}|0\rangle\cdot (i\pmb{\xi}\Delta \tilde{T})+\sum_{\alpha}\Phi^{\alpha}_i\sqrt{\tau_{\alpha}}\tilde{q}^{\alpha}.
\end{equation}
Rearranging the above equation, we obtain a closed expression for $d_i$ and plug it into Eqs.~(\ref{eq:h_expansion}) \& (\ref{eq:define_h}). A closed expression for $\tilde{g}^{\beta}$ is then given by
\begin{equation}\label{eq:NonZeroEgenvectors}
\tilde{g}^{\beta}= -\sqrt{\frac{C_0}{4k_BT_0^2}}\sqrt{\tau_{\beta}}\sum_{\alpha}S^{\alpha\beta}\sqrt{\tau_{\alpha}}\langle \alpha|\mathbf{v}|0\rangle\cdot (i\pmb{\xi}\Delta \tilde{T})+\sqrt{\tau_{\beta}}\sum_{\alpha}S^{\alpha\beta}\sqrt{\tau_{\alpha}}\tilde{q}^{\alpha},
\end{equation}
where the suppression function $S^{\alpha\beta}(\pmb{\xi},\eta)$ is given by 
\begin{equation}
S^{\alpha\beta}(\pmb{\xi},\eta) = \sum_i \frac{\Phi^{\alpha}_i\Phi^{\beta}_i}{1+i\lambda_i(\pmb{\xi},\eta)}.
\end{equation}
Before plugging $\tilde{g}^{\beta}$ into Eq.~(\ref{eq:BTE_zeroth_FourierTransformed}), we first take a closer look at the expression for the heat flux given by
\begin{equation}\label{eq:GeneralizedFourierLaw}
\mathbf{J}(\pmb{\xi},\eta) = -K(\pmb{\xi},\eta)i\pmb{\xi}\Delta \tilde{T}(\pmb{\xi},\eta)+\sum_{\alpha,\beta}M^{\alpha\beta}S^{\alpha\beta}(\pmb{\xi},\eta)\tilde{q}^{\alpha}(\pmb{\xi},\eta).
\end{equation}
In the above expression, $K(\pmb{\xi},\eta)$ is the thermal conductivity tensor given by
\begin{equation}\label{eq:ThermalConductivity}
K^{ij}(\pmb{\xi},\eta) = C_0\sum_{\alpha,\beta}S^{\alpha\beta}(\pmb{\xi},\eta)\sqrt{\tau_\alpha}\langle \alpha|v^{i}|0\rangle\langle0|v^{j}|\beta\rangle\sqrt{\tau_\beta},
\end{equation}
and matrix M is given by
\begin{equation}
M^{\alpha\beta} =\sqrt{4k_BT^2_0C_0} \sqrt{\tau_{\alpha}}\langle0|\mathbf{v}|\beta\rangle\sqrt{\tau_{\beta}}.
\end{equation}
Equation (\ref{eq:GeneralizedFourierLaw}) represents a generalized Fourier's law, valid from ballistic to diffusive regimes. In Ref.~\cite{hua_generalized_2019}, we derived a similar expression under the RTA.  Here, we have extended the concept of the generalized Fourier's law beyond the RTA. There are two parts in Eq.~(\ref{eq:GeneralizedFourierLaw}). The first part represents a convolution between the temperature gradient and space-time-dependent thermal conductivity in real space. The second part is solely determined by the inhomogeneous contribution originating from the boundary conditions and source terms. Similar to the first term, the contribution from the external heat generation to the heat flux is also nonlocal. Using the expression given by Eq.~(\ref{eq:GeneralizedFourierLaw}) and performing an inverse Fourier transform in time and space, the energy conservation equation, Eq.~(\ref{eq:BTE_zeroth}), in real space is then written as 
\begin{equation}\label{eq:MacroGoverningEq}
C_0\frac{\partial\Delta T}{\partial t} -\nabla \cdot \iint K(\mathbf{x}-\mathbf{x'},t-t')\nabla T(\mathbf{x'},t')d\mathbf{x'}dt' = q^{(0)}(\mathbf{x},t)-\nabla \cdot \mathbf{B}(\mathbf{x},t),
\end{equation}
where $\mathbf{B}(\mathbf{x},t)$ is the inverse Fourier transform of $\sum_{\alpha,\beta}M^{\alpha\beta}S^{\alpha\beta}(\pmb{\xi},\eta)\tilde{q}^{\alpha}(\pmb{\xi},\eta)$. Equation (\ref{eq:MacroGoverningEq}) gives the generalized macroscopic governing equation of thermal transport in the absence of boundaries.

\section{Dynamical thermal conductivity}

\begin{figure*}
\centering
\includegraphics[scale = 0.5]{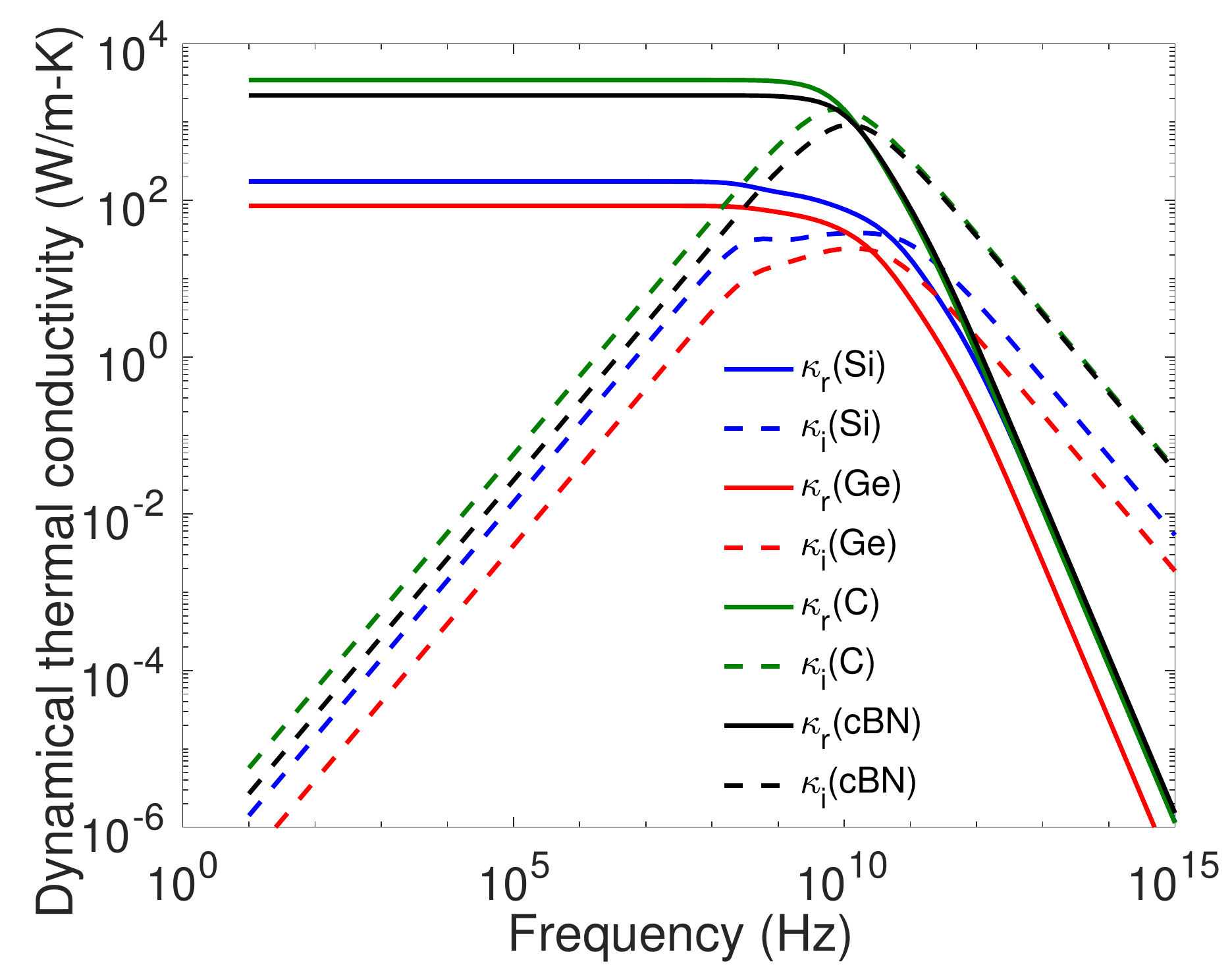}
\caption{The real ($\kappa_r$: solid lines) and imaginary ($\kappa_i$: dashed lines) parts of the dynamical lattice thermal conductivity of diamond, cubic Boron Nitride (cBN), silicon (Si) and germanium (Ge) at room temperature.}
\label{fig:dynamicalthermalconductivity}
\end{figure*}

We first apply this solution in the presence of a constant temperature gradient $\nabla T$. The advective term in Eq.~(\ref{eq:BTE_linearized}) becomes $\mathbf{v_{\mu}}\cdot \nabla n_{\mu} = \mathbf{v_{\mu}} \partial n^0_{\mu}/\partial T \cdot \nabla T$, and Eq.~(\ref{eq:BTE_non-zeroth}) is simplified to
\begin{equation}
\frac{\partial g^{\beta}}{\partial t}+\frac{g^{\beta}}{\tau^{\beta}} = -\sqrt{\frac{C_0}{4k_BT_0^2}}\langle \beta|\mathbf{v}|0\rangle \cdot \nabla(\Delta T)\  \mbox{for } \beta >0. 
\end{equation}
After Fourier transform in time, we derived an analytical solution of the deviational distribution function and a dynamical thermal conductivity as
\begin{equation}
g^{\beta} =- \sqrt{\frac{C_0}{4k_BT_0^2}}\frac{\tau^{\beta}}{1+i\eta\tau^{\beta}}\langle \beta|\mathbf{v}|0\rangle \cdot \nabla(\Delta T),
\end{equation}
and
\begin{equation}\label{eq:dynamicalkappa}
K^{ij}(\eta)= C_0\sum_{\alpha,\beta}\langle \alpha|v^{i}|0\rangle\langle0|v^{j}|\beta\rangle\frac{\tau_\beta}{1+i\eta\tau_\beta}.
\end{equation}
Chaput derived a similar spectral representation for the dynamical thermal conductivity by solving the time-dependent PBE in the irreducible Brillouin zone\cite{chaput_direct_2013}. Here, we demonstrate that our approach is able to reproduce these dynamical thermal conductivities in different materials from first principles methods. 

Four material systems with low (Germanium), medium (Silicon), and high (cubic Boron Nitride and Diamond) thermal conductivities are studied in this work. We use density functional theory (DFT) calculations to determine the phonon dispersions (frequencies and velocities) and anharmonic three-phonon interactions.  However, the formalism presented in this work is applicable to any description of the properties and interactions of the heat carriers, e.g., use of empirical potentials, or inclusion of higher order scatterings or other scattering mechanisms. Specific details of the DFT calculations can be found in Refs. \cite{lindsay_ab_2013} and \cite{lindsay_first-principles_2013}. We used lower integration grids in the first Brillouin zone (60 representative points in the irreducible wedge vs. 408 in Refs. \cite{lindsay_ab_2013} and \cite{lindsay_first-principles_2013}) to determine the phonon scatterings in the example calculations presented here as they are sufficiently converged for our purposes.

The predicted real ($\kappa_r$) and imaginary ($\kappa_i$) parts of the thermal conductivity as a function of temporal frequency are as shown in Fig.~\ref{fig:dynamicalthermalconductivity}. At low frequency, the imaginary part of the thermal conductivity is negligible and the real part remains constant for each material.  The onset of decreased $\kappa_r$ occurs when $\kappa_i$ becomes comparable to $\kappa_r$, which peaks around 0.1 to 1 GHz for different materials. Both $\kappa_r$ and $\kappa_i$ then asymptotically decrease to zero as temporal frequency increases. From Eq.~(\ref{eq:dynamicalkappa}), the temporal effects on thermal conductivity are only observable when the temporal frequency, $\eta$, becomes comparable to the eigenvalues of the collision matrix, $1/\tau_\beta$, which is typically on the order of gigahertz for most solids.

\section{Space-time dependent thermal conductivity in a transient grating experiment}

The size effect on thermal conductivity, on the other hand, is much easier to achieve and observe in an experiment than a pure temporal effect. We now apply our solution method to the geometry of a one-dimensional transient grating (TG) experiment to understand the effects of  geometric length scales on thermal transport. In this experiment, the heat generation rate has a temporal profile of $\delta(t)$ and a spatial profile of e$^{i\rho x}$ in an infinite domain, where $\rho \equiv  2\pi/L$ and $L$ is the grating period. Both the phonon distribution function and temperature field exhibit the same spatial dependence. To simplify the calculation, we assume $\dot{Q}_{\mu}$ is linearly proportional to $\theta^{(0)}_{\mu}$ such that $\tilde{q}^{\alpha} = 0$ for $\forall \alpha >0$. In this assumption, the mode-specific volumetric heat generation, $\hbar\omega_{\mu}\dot{Q}_{\mu}/\text{sinh}(\hbar\omega_{\mu}/2k_BT)$, becomes linearly proportional to the mode-specific volumetric specific heat, which is commonly used in PBE studies\cite{hua_analytical_2014,hua_transport_2014}. The temperature response given by Eq.~(\ref{eq:MacroGoverningEq}) then becomes
\begin{equation}\label{eq:TGTempProfile}
\Delta \tilde{T}(\eta,\rho) = \frac{q^{(0)}}{i\eta C_0+K_{xx}(\eta,\rho)\rho^2}.
\end{equation}
where $K_{xx}(\eta,\rho)$ is the effective thermal conductivity in a 1D TG experiment. The effective thermal conductivity under the RTA for a TG experiment can be found in Ref.~\cite{hua_generalized_2019}. 

When the time scale of a TG experiment is on the order of a few hundred nanoseconds, we can assume that $\sqrt{\tau^{\alpha}}\eta\sqrt{\tau^{\beta}} \ll \sqrt{\tau^{\alpha}}\langle\alpha|\rho v_x|\beta\rangle\sqrt{\tau^{\beta}}$ and $\Gamma^{\alpha\beta} \approx \rho \sqrt{\tau^{\alpha}}\langle\alpha|v_x|\beta\rangle\sqrt{\tau^{\beta}}$ in Eq.~(\ref{eq:BTE_h}) since $\tau_\alpha$ ($\alpha \neq 0$) is typically less than one nanosecond for phonons. While this assumption is valid, the effective thermal conductivity has no temporal dependence, which is consistent with the observation in the dynamical thermal conductivity at low temporal frequency shown in Fig.~\ref{fig:dynamicalthermalconductivity}, and is only a function of grating period, $L$. 
\begin{figure*}
\centering
\includegraphics[scale = 0.42]{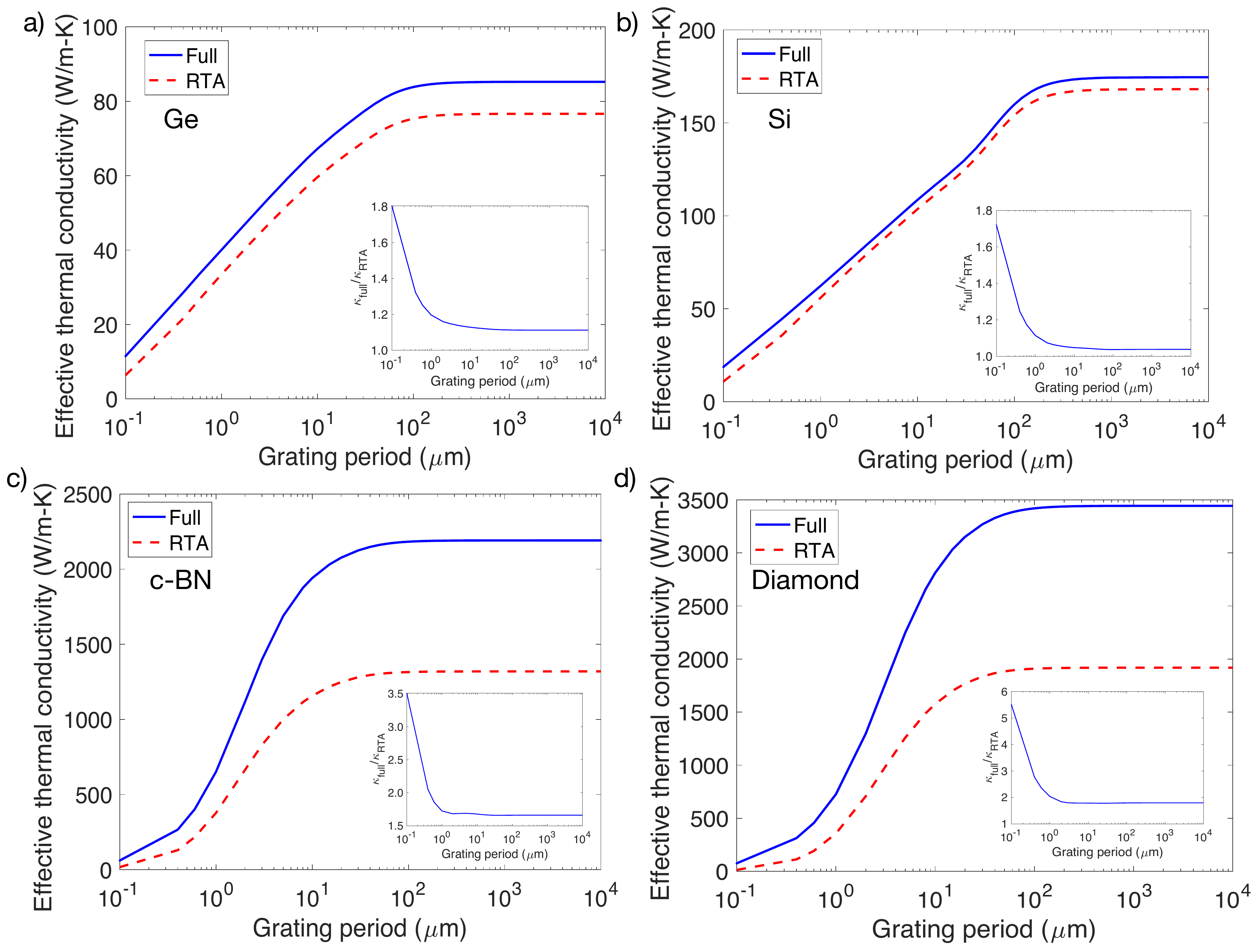}
\caption{Comparison of the effective thermal conductivity as a function of grating period for full (solid lines) and RTA (dashed lines) solutions at room temperature for (a) Ge, (b) Si, (c) c-BN, and (d) diamond. Insets: the ratio between the full and RTA solutions versus grating period. }
\label{Figure1}
\end{figure*}

Figure~\ref{Figure1} gives the effective thermal conductivity as a function of grating period between full and RTA solutions at room temperature for Germanium, Silicon, cubic Boron Nitride (c-BN), and diamond. For diamond and c-BN, the difference in bulk thermal conductivities between the full and RTA solutions is as much as 50 \%, which is consistent with the earlier literature\cite{lindsay_BAs_DFT_2013}. As the grating period decreases, the difference in effective thermal conductivity between the full and RTA solutions increases.  Interestingly, even for Ge and Si where the bulk values given by the RTA and full solutions are within 10 \% difference, the RTA approach fails to give a good estimation of the effective thermal conductivity at smaller grating periods when long MFP phonons are strongly suppressed. This suggests that the actual collision processes are strongly coupled among multiple phonons in nonlocal thermal transport and can not be accurately captured by a single characteristic time. The smallest grating period used in Fig.~\ref{Figure1} is 0.1 $\mu$m, where the onset of temporal dependence is observed in c-BN and diamond at room temperature.

When the time scale of a TG experiment is on the order of a few nanoseconds, $\sqrt{\tau^{\alpha}}\eta\sqrt{\tau^{\beta}} \sim \sqrt{\tau^{\alpha}}\langle\alpha|\rho v_x|\beta\rangle\sqrt{\tau^{\beta}}$. The effective thermal conductivity has both temporal and spatial dependence and must be evaluated using Eq.~(\ref{eq:ThermalConductivity}). The effect of temporal dependence becomes stronger as temperature decreases. To demonstrate this effect, we calculated the effective thermal conductivity dependence on $\eta$ and its corresponding TG temperature responses for c-BN at 100 and 150 K with a grating period of 10 $\mu$m. Note that for this grating period the effective thermal conductivity of c-BN is higher at 150 K.

\begin{figure*}
\centering
\includegraphics[scale = 0.28]{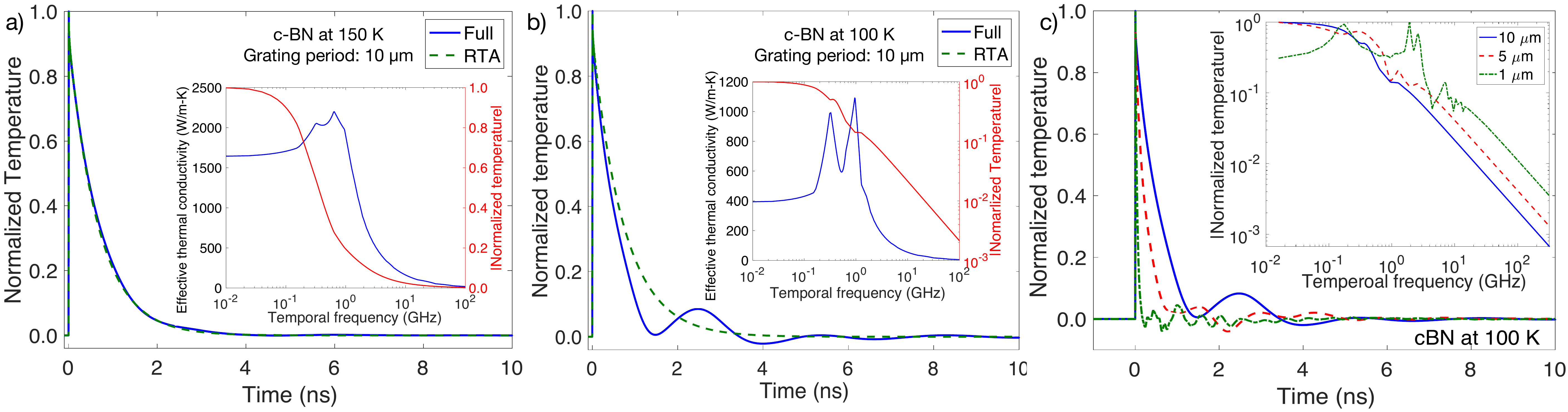}
\caption{Temperature response of c-BN in the time domain at (a) 150 and (b) 100 K with a grating period of 10 $\mu$m predicted by Eq.~(\ref{eq:TGTempProfile}) (solid lines) and a RTA solution given in Ref.~\cite{hua_generalized_2019} (dashed lines). Inset: effective thermal conductivity (left axis) and the magnitude of its corresponding temperature response (right axis) as a function of temporal frequency $\eta$. (c) Time-domain temperature response of c-BN at 100 K with grating periods of 10 (solid line), 5 (dashed lines), and 1 (dotted dashed lines) $\mu$m. Inset: the magnitude of their corresponding temperature responses as a function of temporal frequency $\eta$.}
\label{Figure2} 
\end{figure*}

\begin{figure*}
\centering
\includegraphics[scale = 0.45]{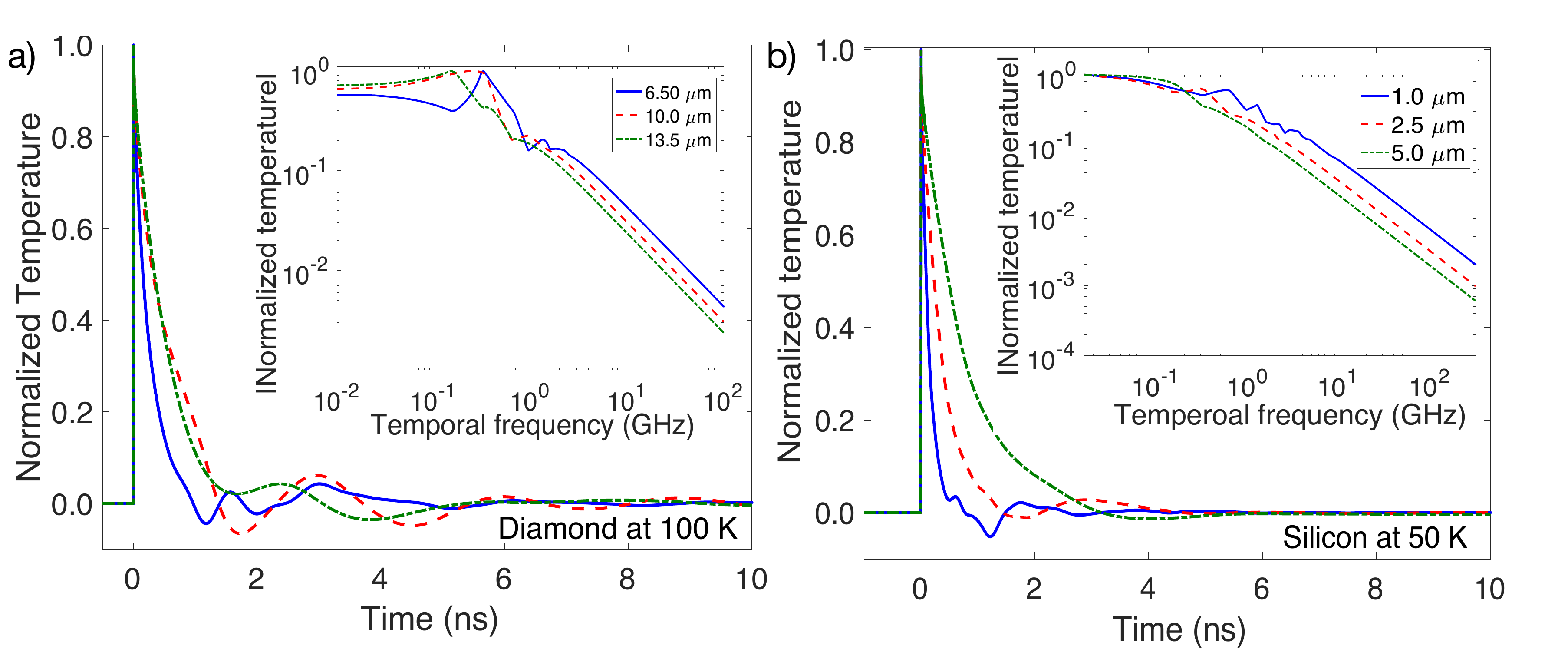}
\caption{Temperature response in time domain of (a) diamond at 100 K and (b) silicon at 50 Kcubic boron nitride with various grating periods. Insets in (a) \& (b): the magnitude of their corresponding temperature responses as a function of temporal frequency $\eta$.}
\label{Figure3}
\end{figure*}

At 150 K, the effective thermal conductivity has a non-monotonic dependence on $\eta$, which peaks around 2 and 4 GHz as shown in the inset of Fig.~\ref{Figure2}(a). Despite the non-monotonic behavior of the effective thermal conductivity, the temperature response in the frequency domain given by Eq.~(\ref{eq:TGTempProfile}) remains a monotonically decreasing function as temporal frequency increases. Therefore, a monotonic decay in the real-space temperature response is observed. At lower temperature (Fig.~\ref{Figure2}(b) inset), the peaks in effective thermal conductivity become more prominent and the temperature frequency response becomes non-monotonic. In real space (as shown in Fig.~\ref{Figure2}), the temperature decays in an oscillatory manner which is absent in the RTA solution. Note that the time scales we consider here are well beyond one nanosecond, and thus the PBE is still valid as it breaks down when the time scales become comparable to the collision time. When the grating period decreases, the oscillation becomes faster as shown in Fig.~\ref{Figure2}(c) and peaks in temperature frequency responses become more prominent as shown in the inset of Fig. \ref{Figure2}(c). 

To further demonstrate that the oscillations due to nonlocal effect are readily observable in other materials, we calculated the TG temperature responses for diamond at 100 K and silicon at 50 K in both time and frequency domains with various grating periods as shown in Fig.~\ref{Figure3}. When the time scale of the temperature response is comparable to the relaxation times of the phonons, thermal conductivity is no longer a constant but depends on both time and space, which itself is due to the nonlocality of the local distribution, $\Delta f_{\mu}$.

A similar oscillatory decay has been observed in graphite above 100 K in transient grating experiments\cite{huberman_observation_2019}, where the oscillation was attributed to the observation of second sound. Here the phase velocities of the observed oscillations in all three materials is much smaller than their theoretical predicted second sound values. The non-trivial relation of nonlocal lattice thermal conductivity and hydrodynamic phonon transport is beyond the scope of this work.

\section{Summary}

To summarize, exploiting the symmetry properties of the linearized collision operator we have derived a direct solution to the space-time-dependent Peierls-Boltzmann transport equation which allows computation of heat flux and temperature fields in nondiffusive regimes without using the relaxation time approximation. The nonlocal thermal transport in a transient grating geometry is studied for the first time in the context of a linearized collision matrix. This also allows for a quantitative estimate of the failure of the RTA approach in describing nonlocal thermal transport. The RTA approach not only fails to predict the bulk thermal conductivity of materials with weak thermal resistance but also fails to accurately characterize the nonlocal effects for materials with relatively low thermal conductivity. Moreover, the thermal conductivity dependence on space and time can change the overall temperature response in nonlocal thermal transport regimes, which could lead to a similar macroscopic response as hydrodynamic transport. The solution method presented here provides the necessary tool to distinguish the two phenomena.  This work provides an accurate mathematical description of spatial and temporal nonlocal thermal transport that will enable critical understanding of materials design for improved thermal management applications.

\section*{Acknowledgments}

The authors gratefully acknowledge A. Minnich for useful discussions. This work was supported by the U. S. Department of Energy, Office of Science, Basic Energy Sciences, Materials Sciences and Engineering Division. 

\appendix
\section*{Appendix: Linearized scattering matrix}

For three-phonon scattering, following the expression given by Ziman\cite{Ziman1960}, the collision operator of the Peierls-Boltzmann equation is written as
\begin{eqnarray}\label{eq:ThreePhononRaw}\nonumber
&&\frac{\partial f_{\mu}}{\partial t}\biggm|_{s} = \sum_{\mu' \mu''}|F_{\mu+\mu'\leftrightarrow\mu''}|^2\delta(-\omega-\omega'+\omega'')\Delta(-\mathbf{q}-\mathbf{q'}+\mathbf{q''})[(f_{\mu}+1)(f_{\mu'}+1)f_{\mu''}-f_{\mu}f_{\mu'}(f_{\mu''}+1)] \\
&&+ \frac{1}{2}\sum_{\mu' \mu''}|F_{\mu\leftrightarrow\mu'+\mu''}|^2\delta(-\omega+\omega'+\omega'')\Delta(-\mathbf{q}+\mathbf{q'}+\mathbf{q''})[(f_{\mu}+1)f_{\mu'}f_{\mu''}-f_{\mu}(f_{\mu'}+1)(f_{\mu''}+1)] 
\end{eqnarray}
where $F_{\mu+\mu'\leftrightarrow\mu''}$ and $F_{\mu\leftrightarrow\mu'+\mu''}$ are the strength of the interaction of the three phonons involved in the scattering\cite{pascual-gutierrez_thermal_2009} and the Kronecker delta $\Delta$ is zero unless its argument is zero or a reciprocal lattice vector, in which case it takes the value $1$. $f_{\mu}$ is the out-of-equilibrium occupation function for all possible phonon states $\mu$ ($\mu \equiv (\mathbf{q},s)$, where $\mathbf{q}$ is the phonon wavevector and $s$ is the phonon polarization). 

Here we define $f_{\mu} \equiv f^0_{\mu} + \Delta f_{\mu}$, where $f^0_{\mu} = (\text{exp}(\hbar\omega_{\mu}/(k_BT_0))-1)^{-1}$ is the global equilibrium Bose-Einsetin distribution. $\omega_{\mu}$ is the phonon frequency, $k_B$ is the Boltzmann constant, and $T_0$ is the equilibrium temperature. Using this definition in Eq.~(\ref{eq:ThreePhononRaw}) and only keeping first order terms in $\Delta f_{\mu}$, Eq.~(\ref{eq:ThreePhononRaw}) becomes
\begin{eqnarray}\nonumber
&&\frac{\partial f_{\mu}}{\partial t}\biggm|_{s} = \sum_{\mu' \mu''}|F_{\mu+\mu'\leftrightarrow\mu''}|^2\delta(-\omega-\omega'+\omega'')\Delta(-\mathbf{q}-\mathbf{q'}+\mathbf{q''})\\
&& \times \left[-\frac{(f^0_{\mu'}+1)f^0_{\mu''}}{f^0_{\mu}}\Delta f_{\mu}-\frac{(f^0_{\mu}+1)f^0_{\mu''}}{f^0_{\mu'}}\Delta f_{\mu'}+\frac{f^0_{\mu}f^0_{\mu'}}{f^0_{\mu''}}\Delta f_{\mu''}\right] \nonumber \\
&&+ \frac{1}{2}\sum_{\mu' \mu''}|F_{\mu\leftrightarrow\mu'+\mu''}|^2\delta(-\omega+\omega'+\omega'')\Delta(-\mathbf{q}+\mathbf{q'}+\mathbf{q''}) \nonumber \\
 &&\times \left[-\frac{f^0_{\mu'}f^0_{\mu''}}{f^0_{\mu}}\Delta f_{\mu}+\frac{f^0_{\mu}(f^0_{\mu''}+1)}{f^0_{\mu'}}\Delta f_{\mu'}+\frac{f^0_{\mu}(f^0_{\mu'}+1)}{f^0_{\mu''}}\Delta f_{\mu''}\right]. \label{eq:LinearizedThreePhonon}
\end{eqnarray}

It is possible to rearrange this linearized scattering operator into a symmetric matrix, $\underline{\underline{\Omega}}$, by defining a new variable $n_{\mu} = \Delta f_{\mu}\text{sinh}(\hbar\omega_{\mu}/2k_BT_0)$ and using the fact that for each collision process ($\mu+\mu'\rightarrow\mu''$), we can find its corresponding decay process ($\mu'' \rightarrow \mu +\mu'$). Therefore, the collision operator can be written as
\begin{equation}
\frac{\partial f_{\mu}}{\partial t}\biggm|_{s} =\frac{1}{\text{sinh}{\frac{\hbar\omega_{\mu}}{2k_BT_0}}}\underline{\underline{\Omega}}\ \underline{n}
\end{equation}
where 
\begin{eqnarray}
\Omega_{\mu\mu} &=& \sum_{\mu'\mu''}|F_{\mu,\mu',\mu''}|^2\delta(-\omega-\omega'+\omega'')\Delta(-\mathbf{q}-\mathbf{q'}+\mathbf{q''})\frac{\text{sinh}{\frac{\hbar\omega_{\mu}}{2k_BT_0}}}{2\text{sinh}{\frac{\hbar\omega_{\mu'}}{2k_BT_0}}\text{sinh}{\frac{\hbar\omega_{\mu''}}{2k_BT_0}}}\\
\Omega_{\mu'\mu'} &=& \sum_{\mu\mu''}|F_{\mu,\mu',\mu''}|^2\delta(-\omega-\omega'+\omega'')\Delta(-\mathbf{q}-\mathbf{q'}+\mathbf{q''})\frac{\text{sinh}{\frac{\hbar\omega_{\mu'}}{2k_BT_0}}}{2\text{sinh}{\frac{\hbar\omega_{\mu}}{2k_BT_0}}\text{sinh}{\frac{\hbar\omega_{\mu''}}{2k_BT_0}}}\\
\Omega_{\mu''\mu''} &=& \sum_{\mu\mu'}|F_{\mu,\mu',\mu''}|^2\delta(-\omega''+\omega+\omega')\Delta(-\mathbf{q}-\mathbf{q'}+\mathbf{q''})\frac{\text{sinh}{\frac{\hbar\omega_{\mu''}}{2k_BT_0}}}{2\text{sinh}{\frac{\hbar\omega_{\mu}}{2k_BT_0}}\text{sinh}{\frac{\hbar\omega_{\mu'}}{2k_BT_0}}}\\
\Omega_{\mu\mu'} &=& \Omega_{\mu'\mu} = \sum_{\mu''}|F_{\mu,\mu',\mu''}|^2\delta(-\omega-\omega'+\omega'')\Delta(-\mathbf{q}-\mathbf{q'}+\mathbf{q''})\frac{1}{2\text{sinh}{\frac{\hbar\omega_{\mu''}}{2k_BT_0}}}\\
\Omega_{\mu\mu''} &=& \Omega_{\mu''\mu} = \sum_{\mu'}|F_{\mu,\mu',\mu''}|^2\delta(-\omega-\omega'+\omega'')\Delta(-\mathbf{q}-\mathbf{q'}+\mathbf{q''})\frac{1}{2\text{sinh}{\frac{\hbar\omega_{\mu'}}{2k_BT_0}}}\\
\Omega_{\mu'\mu''} &=& \Omega_{\mu''\mu'} = \sum_{\mu}|F_{\mu,\mu',\mu''}|^2\delta(-\omega-\omega'+\omega'')\Delta(-\mathbf{q}-\mathbf{q'}+\mathbf{q''})\frac{1}{2\text{sinh}{\frac{\hbar\omega_{\mu}}{2k_BT_0}}}.
\end{eqnarray}

\bibliographystyle{unsrt}
\bibliography{myref}

\end{document}